\documentclass[conference, 10pt]{IEEEtran}

\usepackage{cite}
\usepackage{url}
\usepackage{graphicx}
\usepackage{color}
\usepackage{placeins}
\usepackage{float}
\usepackage{tabularx,colortbl}
\usepackage{ifthen}
\usepackage{circuitikz}
\usepackage{multirow}
\usepackage{subcaption}
\usepackage{pgfplots}
\usepackage{tikz}
\usepackage{environ}
\makeatletter
\newsavebox{\measure@tikzpicture}
\NewEnviron{scaletikzpicturetowidth}[1]{%
  \def\tikz@width{#1}%
  \begin{lrbox}{\measure@tikzpicture}%
  \BODY
  \end{lrbox}%
  \pgfmathparse{#1/\wd\measure@tikzpicture}%
  \BODY
}
\usepackage[linesnumbered,ruled,lined]{algorithm2e}

\usepackage[colorlinks=true, allcolors=blue]{hyperref}

\makeatletter

\newcounter{author}
\renewcommand{\author}[2][]{
   \stepcounter{author}
   \@namedef{author@\theauthor}{#2}
   \@namedef{authorlabel@\theauthor}{#1}
}

\newcounter{address}
\newcommand{\address}[2][]{
   \stepcounter{address}
   \@namedef{address@\theaddress}{#2}
   \@namedef{addresslabel@\theaddress}{#1}
}

\newcommand{\alsep}{and}

\def\newmaketitle{\par%
  \begingroup%
  \normalfont%
  \def\thefootnote{}
  \def\footnotemark{}
  \let\@makefnmark\relax
  \footnotesize
  \footnotesep 0.7\baselineskip
  \normalsize%
  \twocolumn[\thenewmaketitle\@IEEEaftertitletext]%
  \if@IEEEusingpubid
     \enlargethispage{-\@IEEEpubidpullup}%
  \fi
  \endgroup
  \setcounter{footnote}{0}\let\maketitle\relax\let\@maketitle\relax
  \gdef\@thanks{}%
  \let\thanks\relax}

\def\thenewmaketitle{
  \newpage
  \begin{center}%
    \vskip0.2em{\Huge\@IEEEcompsoconly{\sffamily}\@IEEEcompsocconfonly{\normalfont\normalsize\vskip 2\@IEEEnormalsizeunitybaselineskip
   \bfseries\large}\@title\par}\vskip1.0em\par%
    \vspace{1ex}
    \newcounter{c@author}
    \newcounter{c@tmp}
    \ifthenelse{\value{author}=2}{%
      \newcommand{\liand}{ and }}{%
      \newcommand{\liand}{, and }}
    \ifthenelse{\value{address}<2}{%
      \@nameuse{author@1}%
      \stepcounter{c@author}%
      \whiledo{\value{c@author}<\value{author}}{%
        \setcounter{c@tmp}{\value{author}}%
        \addtocounter{c@tmp}{-\value{c@author}}%
        \ifthenelse{\value{c@tmp}=1}{%
          \renewcommand{\alsep}{\liand}}{\renewcommand{\alsep}{, }}%
        \stepcounter{c@author}\alsep \@nameuse{author@\thec@author}}\\%
    }
    {
      \@nameuse{author@1}${}^{(\ref{\@nameuse{authorlabel@1}})}$%
      \stepcounter{c@author}%
      \whiledo{\value{c@author}<\value{author}}{%
      \setcounter{c@tmp}{\value{author}}%
      \addtocounter{c@tmp}{-\value{c@author}}%
      \ifthenelse{\value{c@tmp}=1}{%
        \renewcommand{\alsep}{\liand}}{\renewcommand{\alsep}{, }}%
      \stepcounter{c@author}\alsep \@nameuse{author@\thec@author}%
        ${}^{(\ref{\@nameuse{authorlabel@\thec@author}})}$%
      }
    }
    \vspace{0.2ex}

    \ifthenelse{\value{address}>0}{%
      \ifthenelse{\value{address}=1}{
        {\@nameuse{address@1}}
      }
      {
        \newcounter{c@address}

        \begin{center}
        \whiledo{\value{c@address}<\value{address}}
        {
          \refstepcounter{c@address}
            ${}^{(\thec@address)}$\,%
              \label{\@nameuse{addresslabel@\thec@address}}%
              \@nameuse{address@\thec@address}\\ %
        }
        \end{center}
      } 
    }
    {
      \relax
    }
  \end{center}
}

\makeatother

\title{A Tunable Reflection Surface with Independently Variable Phase and Slope}

\author[org1]{Omran Abbas,~\IEEEmembership{Student~Member,~IEEE,}
Anas Chaaban,~\IEEEmembership{Senior Member,~IEEE,}
and Lo\"{i}c~Markley,~\IEEEmembership{Senior Member,~IEEE}}

\address[org1]{The University of British Columbia, Kelowna, Canada (e-mail: omran.abbas, anas.chaaban, loic.markley@ubc.ca).}

\begin{document}

\newmaketitle

\begin{abstract}
A reconfigurable intelligent surface (RIS) is an essential component in the architecture of the next generation of wireless communication systems. An RIS is deployed to provide a \textcolor{black}{controllability} to the multi-path environment between the transmitter and the receiver, which becomes critical when the line-of-sight signal between them is blocked.
In this work, we design an electrically tunable linearly polarized RIS at $2.5$~GHz that yields a controllable reflection phase and phase-frequency slope; in other words, we add tunability of the phase-frequency slope to the tunability of the resonance center frequency. The proposed design consists of two layers of unit cells \textcolor{black}{placed over a ground plane, with} dog-bone-shaped elements in the top layer and patch elements in the bottom layer. Each patch and dog-bone element is loaded with a varactor, whose reverse bias voltage is controlled to provide a phase-frequency profile with a slope value of $9$~degrees/MHz or $0.95$~degrees/MHz, and a phase shift range of $320$ degrees.
\end{abstract}

\vspace{-0.25cm}
\section{Introduction}
A Reconfigurable intelligent surface (RIS) works as a controllable spatial filter \cite{aless} that can be optimized along with the communication endpoints to bring about a higher coverage and a better quality of service.
The RIS, in principle, is an array of unit cells whose reflection phase can be independently controlled to create a phase gradient across the RIS surface, which generates a beam toward a specific direction \cite{SRE1}. Typically, an RIS does not employ a complex phase shifters circuit \cite{RAAdv}, which decreases its implementation cost and complexity. The unit cell is characterized by a resonance curve describing the amplitude and the phase response across a certain frequency band. The controllability over the resonance profile can be realized using mechanical control, to physically adjust the array \cite{RAMech}, or electrical control with lumped elements \textcolor{black}{like} varactors \cite{SeanRIS} or PIN \cite{PINRA} \textcolor{black}{that load the unit cells}.

\textcolor{black}{Current unit cell designs offer} phase shift tunability over wide bands, with a small change in the slope of the phase-frequency profile. Such designs are well suited for narrow-band wireless communication systems where a constant phase-frequency response is desired \cite{RISCommUsual}. However, the results presented in \cite{RISMultiSlope} for a wideband system showed the advantage of \textcolor{black}{ deploying an RIS with unit cells that have a} phase-frequency profile with both tunable slopes and phase shifts, especially when large arrays are considered. \textcolor{black}{In this work, we} propose a unit cell structure that has a resonance profile with two different slopes over a $200$~MHz band centered around $2.5$~GHz, \textcolor{black}{while maintaining a tunable phase shift.}

\vspace{-0.25cm}
\section{Unit Cell Design} \label{operation}

\begin{figure}[!t]
\centering
\tikzset{every picture/.style={scale=.6}, every node/.style={scale=.9}}
\input{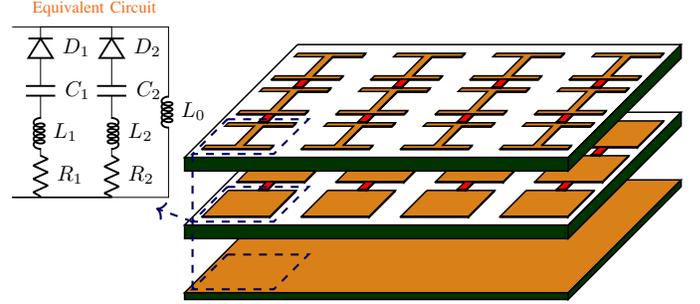}
\caption{An expanded view of the two-slope RIS design with parallel unit cells having dog-bone-shaped elements in the first layer and patch elements in the second layer, placed over the ground plane. The red rectangles are the varactors, the orange areas are copper, and the white areas are the insulating material. The equivalent circuit of the parallel unit cell is also shown.}
\label{RISCircuitr}
\vspace{-0.25cm}
\end{figure}

The proposed linearly-polarized array design and the approximate equivalent circuit of each unit cell are shown in Fig. \ref{RISCircuitr}. \textcolor{black}{The array consists of two layers of varactor-loaded elements placed over a ground plane. The top layer is comprised of dog-bone-shaped elements while the middle layer is comprised of patch elements.} 
The dimensions of the elements and the reverse bias voltage applied to the varactors determine the characteristics of the resonance profile of the unit cell. Specifically, by changing the varactor biasing, we change the equivalent capacitance of each element. \textcolor{black}{This allows us to electrically control the characteristics of the unit cell resonance curve, specifically, the reflection phase and the phase-frequency slope at the operating frequency.}

\begin{figure}[!h]
\centering
\tikzset{every picture/.style={scale=.6}, every node/.style={scale=.9}}
\input{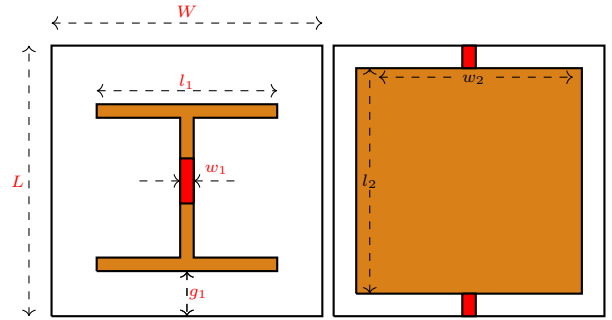}
\caption{The Configurations of the dog-bone and patch elements. The red rectangles are the positions of the varactors in the two elements. The dimensions considered in simulations are $L = 27$~mm, $W = 27$~mm, $l_{1} = 10$~mm, $w_{1} = 0.5$~mm, $g_{1} = 7$~mm, $l_{2} = 26$~mm, and $w_{2} = 20$~mm. The substrate is Rogers 5880 with a thickness of $1.57$~mm.}
\label{UnitCell}
\vspace{-0.25cm}
\end{figure}

The dog-bone-shaped element can be \textcolor{black}{modeled by} an RLC series resonator with a resistance $R_{1}$, an inductance $L_{1}$, and a capacitance $C_{1}$ connected in series with a varactor $D_{1}$. Similarly, the patch element can be \textcolor{black}{modelled by} an RLC series resonator with a resistance $R_{2}$, an inductance $L_{2}$, and a capacitance $C_{2}$ connected in parallel with a varactor $D_{2}$. The ground is modelled as an inductor $L_{0}$. The equivalent circuit of the unit cell is an inductance in parallel with the two series resonators. By controlling the resonance frequency of the resonators, a different slope of the phase-frequency profile can be realized.

The resonance profile of the parallel unit cell is categorized into two domains: the dog-bone domain (maximum slope $S_{max}$) and the patch domain (minimum slope $S_{min}$). Each domain will have a single slope and different phase shift values. 
The first domain provides a resonance profile with the maximum phase-frequency slope by applying a fixed bias voltage on $D_{2}$ and \textcolor{black}{varying} bias voltages on $D_1$, resulting in the resonance profile of the dog-bone-shaped element. The voltage applied on $D_{2}$ causes the patch layer to \textcolor{black}{effectively} function as a reflector, maintaining a consistent phase shift, while the voltage on $D_1$ keeps the resonance frequency of the dog-bone element within the band of interest.
On the other hand, to get the minimum phase-frequency slope, a variable bias voltage is applied on $D_{2}$ \textcolor{black}{to make} the patch resonate in the band of interest. A bias voltage \textcolor{black}{is then applied on} $D_1$ to make the dog-bone-shaped element resonate \textcolor{black}{away from the operating frequency.}

\vspace{-0.25cm}
\section{Simulation Results} \label{simuation}

\begin{figure}[!t]
   \begin{minipage}{0.48\textwidth}
     \centering
     \tikzset{every picture/.style={scale=.67}, every node/.style={scale=.9}}
    \input{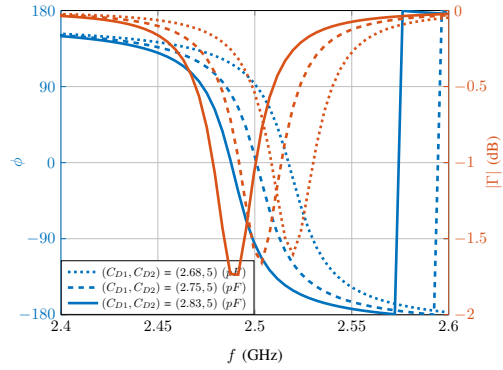}
    \subcaption{Dog-bone domain where $C_{D2}$ is fixed to $5$~pF while $C_{D1}$ is being changed between $2.68$~pF and $2.83$~pF.}
   \end{minipage}\hfill
   \begin{minipage}{0.48\textwidth}
     \centering
     \tikzset{every picture/.style={scale=.67}, every node/.style={scale=.9}}
    \input{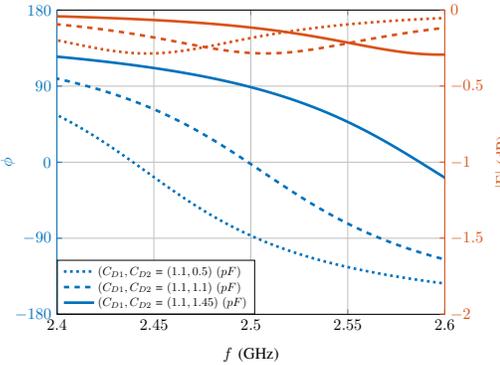}
    \subcaption{Patch domain where $C_{D1}$ is fixed to $1.1$~pF while $C_{D2}$ is being changed between $0.5$~pF and $1.45$~pF.}
   \end{minipage}
   \caption{The Amplitude-, in orange, and the phase-frequency, in blue, responses of the unit cell shown in \ref{UnitCell}, where $C_{D1}$ and $C_{D2}$ are the equivalent capacitance of the diode on the top and bottom layer, respectively.}
\label{ParallelUnitCellr}
\vspace{-0.25cm}
\end{figure}

The resonance profile of the unit cell is simulated using ANSYS High-Frequency Simulation Software and plotted in Fig. \ref{ParallelUnitCellr}.
The resonance profile of the unit cell shows two slopes of the phase-frequency profile \textcolor{black}{over three different phase shifts. These profiles are associated with different equivalent varactor capacitances, labelled as $C_{D1}$ and $C_{D2}$.}
Specifically, Fig. \ref{ParallelUnitCellr}(a) presents the case where the slope of the phase-frequency profile is $S_{max} = 9$~degrees/MHz, and the phase shift varies from $-90$~degrees for $(C_{D1}, C_{D2}) =(2.83,5)$~pF to $90$~degrees for $(C_{D1}, C_{D2}) =(2.68,5)$~pF. Fig. \ref{ParallelUnitCellr}(b) shows the case where the slope of the phase-frequency profile is $S_{min} = 0.9$~degrees/MHz, and the phase shift varies from $-90$ to $90$ degrees. \textcolor{black}{It is worth mentioning that we used the SMV2023 data sheet to model the varactors in the simulation results shown in Fig. \ref{ParallelUnitCellr}}

We also observe that the reflection loss increases as the phase-frequency slope increases. Specifically, it increases from $0.3$~dB at $S_{min}$ to $2$~dB at $S_{max}$. The losses are primarily due to the finite conductivity of the copper traces and the varactor resistance, \textcolor{black}{which increases for the} dog-bone-shaped elements. 

\vspace{-0.25cm}
\section{Conclusion} \label{conclusion}

This paper proposed an RIS unit cell design in which the slope and the phase shift of the phase-frequency profile could be controlled. The results have shown a phase-frequency profile with a slope value of $9$~degrees/MHz or $0.95$~degrees/MHz and multiple phase shift values between $-160$~degrees and $160$~degrees for both slopes.

\end{document}